\documentclass[preprint,aps]{revtex4}

\usepackage{graphicx}

\begin{document}

\title{Electronic Structure of Transition Metal Dichalcogenides PdTe$_2$ and Cu$_{0.05}$PdTe$_2$ Superconductors Obtained by Angle-Resolved Photoemission Spectroscopy}

\author{Yan Liu$^{1}$, Jianzhou Zhao$^{1}$, Li Yu$^{1}$, Chengtian Lin$^{2}$, Cheng Hu$^{1}$, Defa Liu$^{1}$, Yingying Peng$^{1}$, Zhuojin Xie$^{1}$, Junfeng He$^{1}$, Chaoyu Chen$^{1}$, Ya Feng$^{1}$, Hemian Yi$^{1}$, Xu Liu$^{1}$, Lin Zhao$^{1}$, Shaolong He$^{1}$, Guodong Liu$^{1}$, Xiaoli Dong$^{1}$, Jun Zhang$^{1}$, Chuangtian Chen$^{3}$, Zuyan Xu$^{3}$, Hongming Weng$^{1}$, Xi Dai$^{1}$, Zhong Fang$^{1}$ and X. J. Zhou$^{1,4,*}$
}

\affiliation{
\\$^{1}$National Laboratory for Superconductivity, Beijing National Laboratory for Condensed Matter Physics, Institute of Physics, Chinese Academy of Sciences, Beijing 100190, China
\\$^{2}$Max-Planck-Institut f\"{u}r Festk\"{o}rperforschung, Heisenbergstrasse 1, 70569 Stuttgart, Germany
\\$^{3}$Technical Institute of Physics and Chemistry, Chinese Academy of Sciences, Beijing 100190, China
\\$^{4}$Collaborative Innovation Center of Quantum Matter, Beijing 100871, China
\\$^{*}$Corresponding authors: XJZhou@aphy.iphy.ac.cn
}
\date{May 25, 2015}

\begin{abstract}
The layered transition metal chalcogenides  have been a fertile land in solid state physics for many decades. Various $MX_2$-type transition metal dichalcogenides, such as WTe$_2$, IrTe$_2$, and MoS$_2$, have triggered great attention recently, either for the discovery of novel phenomena or some  extreme or exotic physical properties, or for their potential applications. PdTe$_2$ is a superconductor in the class of transition metal dichalcogenides, and superconductivity is enhanced in its Cu-intercalated form, Cu$_{0.05}$PdTe$_2$. It is important to study the electronic structures of PdTe$_2$ and its intercalated form in order to explore for new phenomena and physical properties and understand the related superconductivity enhancement mechanism. Here we report systematic high resolution angle-resolved photoemission (ARPES) studies on PdTe$_2$  and Cu$_{0.05}$PdTe$_2$ single crystals, combined with the band structure calculations. We present for the first time in detail the complex multi-band Fermi surface topology and densely-arranged band structure of these compounds. By carefully examining the electronic structures of the two systems, we find that Cu-intercalation in PdTe$_2$ results in electron-doping, which causes the band structure to shift downwards by nearly 16 meV in Cu$_{0.05}$PdTe$_2$. Our results lay a foundation for further exploration and investigation on PdTe$_2$ and related superconductors.
\end{abstract}

\maketitle

\section{Introduction}

The transition metal chalcogenides have been a fertile land for studying new phenomena and properties in condensed matter physics, including the traditional study of charge density wave (CDW) and related superconductivity\cite{CDW,EMorosan} and the latest discovery of three-dimensional topological insulators.\cite{YXia,HJZhang,YLChen}  Among them, a number of $MX_2$-type transition metal dichalcogenides, including IrTe$_2$, WTe$_2$, MoS$_2$, and MoSe$_2$, have attracted great attention recently due to their rich physical properties like charge-density wave,\cite{McCarron,Wilson1975,Moncton} superconductivity,\cite{Smaalen,Wilson1969,Morris,Finlayson,Sipos} catalysis of chemical reactions,\cite{YLi} pressure-induced structural rearrangement,\cite{Soulard} extremely large magneto-resistance\cite{Mazhar,PLCai} and potential technological applications.\cite{Mazhar,HYLv,Yeh,PanXC,Duerloo,YiZhang} There is still plenty of room to explore for new phenomena and physical properties in the transition metal dichalcogenides, particularly in those systems that have been paid less attention before, such as the systems consisting of the VIII metal  and chalcogen atoms such as PdTe$_2$, NiTe$_2$, PtSe$_2$, PtS$_2$, and so on.  The Pd--Te system can form various intermetallic compounds like PdTe$_2$, PdTe, Pd$_3$Te$_2$, and Pd$_4$Te.\cite{Tibbals,Thomassen,Matthias5390,Matthias5392,Gronvold,Kjekshus1959,Westrum,Medvedeva,Kjekshus1965,Furuseth,Ipser,CMallika,Simic}
Among them, the most studied CdI$_2$-type PdTe$_2$ and NiAs-type PdTe are superconductors with  $T_c$ of $1.7$~K\cite{Finlayson,Roberts} and  $2.3$--$4.5$ K,\cite{Karki,Matthias5390,Matthias5392,Kjekshus1965} respectively.  Understanding the electronic structures of these compounds is important for understanding the related superconductivity mechanism and exploring for new phenomena and physical properties.

In this paper, we focus on the investigation of the electronic structures of PdTe$_2$ and its Cu-intercalated counterpart Cu$_{0.05}$PdTe$_2$.  Few work has been reported in the literature except for some earlier scanning tunneling spectroscopy\cite{Ryan1999,Ryan2000} and angle-resolved photoemission studies on PdTe$_2$.\cite{Orders}  We carried out systematic high resolution angle-resolved photoemission (ARPES) measurements on the electronic structures of PdTe$_2$ and Cu$_{0.05}$PdTe$_2$. Multiple bands with strong dispersions are revealed over a wide energy range (0--6 eV), which give rise to a complex Fermi surface topology.  We also performed detailed band structure calculations for PdTe$_2$ by considering the spin--orbit coupling effect. The calculated band structure exhibits strong three-dimensionality. The comparison with the theoretical calculations facilitates the assignment of the observed band structure and Fermi surface, with some observed features not accounted for in the bulk band calculations. While the overall electronic structure is similar between PdTe$_2$ and Cu$_{0.05}$PdTe$_2$, some delicate differences are identified. In particular, there is a downshift of band on the order of 16 meV observed in  Cu$_{0.05}$PdTe$_2$, indicating the electron doping from Cu-intercalation in the system.

\section{Methods}
High quality single crystals of Cu$_x$PdTe$_2$ ($x=0$ and 0.05) were obtained by the self-flux method. The ARPES measurements were performed on our lab photoemission system equipped with the Scienta R4000 electron energy analyzer.\cite{GDLiu} We used two kinds of photon sources for the ARPES measurements. One is the the helium discharge lamp,  which provides helium I line with a photon energy of $h\nu=21.218$ eV. The overall energy resolution used is 10--20 meV and the angular resolution is $\sim$0.3$^{\circ}$.  The other source is a vacuum ultra-violet (VUV) laser with a photon energy of 6.994 eV.\cite{GDLiu} The overall energy resolution used in this case is 1.5 meV and the angular resolution is $\sim$0.3$^{\circ}$ (corresponding to a momentum resolution of  $\sim$0.004 {\AA}$^{-1}$).  The overall electronic structure over large energy scale and momentum space was measured by using the helium lamp, while some fine and high resolution measurements were carried out by using the VUV laser.  The Fermi level was referenced by measuring the Fermi edge of a clean polycrystalline gold electrically connected to the sample. The orientation of the crystals was determined by Laue diffraction at room temperature. The crystals were cleaved {\it in situ} and measured at a temperature of $T\sim 20$ K in vacuum with a base pressure better than 5$\times$10$^{-11}$ Torr.

The electronic structure of PdTe$_2$ was calculated by employing the generalized gradient approximation (GGA) for the electron correlations.  We used the code WIEN2k based on the full-potential linearized augmented-plane-wave method.\cite{Blaha}  The spin--orbit coupling (SOC) was included self-consistently in the ${\rm GGA+SOC}$ calculation. The lattice constants we used were $a=b=4.036$ {\AA} and $c=5.13$ {\AA} for PdTe$_2$, taking a space group of $P\overline{3}m1$.\cite{McCarron,Finlayson,SJobic,WSKim} The Brillouin zone integration was performed on a regular mesh of $12\times 12\times 6$ $k$ points. The muffin-tin radii ($R_{MT}$) of 2.50 bohr were chosen for both Pd and Te atoms. The largest plane-wave vector $K_{max}$ was given by $R_{MT}K_{max}=7.0$. Our calculation results are consistent with the previous reports.\cite{Myron,GYGuo,JPJan}

\section{Results and discussion}
Figure 1 shows the crystal structure and the corresponding Brillouin zone of the parent compound PdTe$_2$£¬ as well as the characterization information of the intercalated Cu$_{0.05}$PdTe$_2$ single crystal. The crystal structure of PdTe$_2$ (Fig. 1(a)) is polymeric CdI$_2$-type with the $P\bar{3}m1$ (No. 164) space group.\cite{McCarron,Soulard,SJobic,WSKim,JPJan} The palladium atom is octahedrally coordinated by six tellurium atoms. Each Pd--Te layer consists of a two-dimensional close-packed array of palladium atoms sandwiched between two similar arrays of tellurium atoms.\cite{Wilson1969,Finlayson,GYGuo} Within the layer, the atoms are covalently bonded, while between the layers, they are bonded weakly in the van der Waals style. Thus the crystal can be easily cleaved between the layers. The corresponding Brillouin zone is a hexagonal prism as shown in Fig. 1(b). Magnetization measurement of the Cu$_{0.05}$PdTe$_2$ crystal was done  in the field-cooled (FC) and zero field-cooled (ZFC) modes with a magnetic field of 1 Oe, giving a superconducting transition temperature $T_c\sim2.4$ K (Fig. 1(c)), which is obviously higher than that of the parent compound PdTe$_2$ ($T_c\sim 1.7$ K).\cite{Finlayson,Roberts} The dimension of the single crystal is about 5 mm$\times$3 mm$\times$0.2 mm, with a shining surface as shown in the inset of Fig. 1(c). Single crystal x-ray diffraction was carried out on the Cu$_{0.05}$PdTe$_2$ sample (Fig. 1(d)). The observed peaks can be indexed into (0 0 $n$)(with $n$ being integers), indicating that the naturally cleaved surface is the basal plane. Based on the Laue diffraction pattern shown in Fig. 1(e), we can determine the orientation of the crystal. The low energy electron diffraction (LEED) pattern on the cleaved surface (Fig. 1f) taken at 25~K at the electron energy of 70 eV is shown in Fig. 1(f).  It is consistent with our Laue measurement. The pattern of sharp and bright spots  demonstrates a three-fold symmetry; no surface reconstruction is observed.

Figure 2 shows the constant energy contours of the ARPES spectral weight distribution for both PdTe$_2$ (Fig. 2(a)) and Cu$_{0.05}$PdTe$_2$ (Fig. 2(b)) at different binding energies. Overall speaking, the constant energy contours of PdTe$_2$ and Cu$_{0.05}$PdTe$_2$ samples are quite similar; they both evolve strongly with the binding energy. Near the Fermi level, both compounds exhibit a complex Fermi surface topology with a three-fold symmetry.  A couple of main features can be well resolved.  A small electron-like spot appears around the $\Gamma$ point, denoted as $\alpha$. It is surrounded by six small electron-like spots ($\beta$ and $\beta'$) located near the midpoint between $\Gamma$ and $K$. There is a rather weak spectral weight that links in between the $\beta$ and $\beta'$ spots.  There is also a weak spectral weight extending out from each $\beta$ spot to the $M$ and $M'$ points, marked as $\delta$ and $\delta'$ bands, respectively.  These observations show some resemblance to that found in the parent IrTe$_2$\cite{Ootsuki2013} and doped Ir$_{0.95}$Pt$_{0.05}$Te$_2$\cite{Ootsuki2014JPSJ} single crystals. Although in that case, the six inner hole pockets are six-fold symmetric. Right outside the first Brillouin zone boundary, there  appear two flap-like sheets (denoted as $\gamma$ and $\gamma'$) popping out symmetrically with respect to the  $\Gamma$--$K$ and $\Gamma$--$K'$ lines, respectively. They also exhibit the three-fold symmetry. Going further into the second zone,  more complicated features appear.  With the increasing binding energy, the $\alpha$ spot near the $\Gamma$ point shrinks, while the three $\beta$ and three $\beta'$ spots near the $\Gamma$--$K$ and $\Gamma$--$K'$ middle points first diminish quickly and then reemerge gradually to become warped square pockets (Figs. 2(a2)--2(a5) and Figs. 2(b2)-2(b5)). Near the $K$ ($K'$) regions, the two flap-like $\gamma$ features at the Fermi level move away from the K (K$'$) points and disappear quickly with increasing binding energy while another two big pockets approach to the K (K$'$) points (Figs. 2(a4) and 2(a5) and Figs. 2(b4) and 2(b5)).  We note that, in both compounds, the measured electronic structure in the second Brillouin zone does not reproduce that in the first Brillouin zone. This may be caused by the strong momentum dependent photoemission matrix-element effect that can alter the relative spectral intensities of various bands. It is also possible that this is due to the strong three-dimensionality of the measured materials, i.e., the band structures show strong k$_z$ dependence, as we will see below from the band structure calculations. For ARPES with a given photon energy of 21.2 eV, the covered momentum space is a curved surface with different k$_z$, which can be obviously different between the first and the second Brillouin zones.

Figure 3 shows the overall band structure of Cu$_{0.05}$PdTe$_2$ measured along three high symmetry momentum cuts with their locations marked in Fig. 3(a). Figures 3(b)--3(d) show the original data, while figures 3(e)-3(g) show the corresponding second-derivative images with respect to energy in order to better highlight some bands.  Multiple dispersive bands are observed within a large energy range up to $\sim$6 eV binding energy that we have covered. These results are consistent with the earlier theoretical calculations\cite{GYGuo,JPJan} and the ARPES work on PdTe$_2$\cite{Orders} that suggested nine hybridized bands composed of Pd 4d-states and Te 5p-states. Our new high-resolution ARPES data shown here undoubtedly provide clearer and comprehensive electronic structure information.  From Figs. 3(b) and 3(c), near the Fermi level, an electron-like band can be seen at the $\Gamma$ point, which corresponds to the $\alpha$-Fermi pocket near $\Gamma$ in Fig. 3(a). In addition, near the midpoint of $\Gamma$--$K$ ($K'$), small electron-like $\beta$ ($\beta'$) bands cross the Fermi level. The subtle difference in the band structures near the Fermi level between $\Gamma$--$K$ (Figs. 3(b) and 3(e)) and $\Gamma$--$M$ (Figs. 3(c) and 3(f)) accounts for the difference in the Fermi surface topology and the constant energy contours (Figs. 2 and 3(a)) along the two momentum directions. At high binding energies (above 1 eV), the measured band structures along $\Gamma$--$M$ and $\Gamma$--$K$ look similar qualitatively.

Figure 4 focuses on the low-energy electronic structures of Cu$_{0.05}$PdTe$_2$ and shows the detailed momentum evolution of the band structures, which helps to understand the features in the measured Fermi surface.  Along the representative momentum cut 1, the measured band structure (Fig. 4(b), cut 1) shows four main features: $\alpha$ band near the $\Gamma$ point,  $\beta$ and $\beta$$'$ bands sitting near the middle of $\Gamma$--$K$ and $\Gamma$--$K'$, respectively, and $\gamma$ band just outside of the $K$ point.   The $\alpha$ band shows an electron-like parabolic dispersion for cut 1; the band becomes narrower for cut 2 and disappears for cut 3 in Fig. 4(b), indicating a small electron pocket near the $\Gamma$ point in the measured Fermi surface (Fig. 4(a)). The most prominent $\beta$ bright spots in the measured Fermi surface (Fig. 4a) appear to be  singular spots, because they have  dramatically high intensities  and are  localized in a rather small momentum space. Right beneath the $\beta$ spot,  one can identify some narrow and relatively deep hole-like bands. These hole-like bands are responsible for the warped square-like pocket near the same location at higher binding energy, as shown in Figs. 2(a) and 2(b). They are localized in the area around the $\Gamma$--$K$ middle points. Even though the $\beta'$ features show rather weak spectral intensity compared with the $\beta$ bands, the corresponding complex electronic behaviors are quite similar.

Furthermore, several features can grow out from these $\beta$ and $\beta$$'$ pockets. First, two groups of bands can extend out from each $\beta$ (but not $\beta$$'$) pocket towards the adjacent M and M$'$ points on the zone boundary, symmetric with respect to the $\Gamma$--$K'$ lines,  marked as $\delta$ and $\delta$$'$ bands in Figs. 2(b) and  4(a).  These features further connect with some complicated structures in the second Brillouin zone.  The evolution of the corresponding bands in Fig. 4(b) (cuts 1--5) clearly indicates that these $\delta$ ($\delta'$) bands are electron-like. From Fig. 4(b), the electron-like $\delta$ bands seem to grow naturally from the $\beta$ bands in terms of band structure evolution. However, the singular behavior of the particularly strong intensity of the $\beta$ bright spots seems not to be compatible with this evolution. Whether it is the same as the $\delta$ band or it comes from a different origin needs to be further investigated. Second, there are also some weak features that link all the $\beta$ and $\beta$$'$ points together to form a hexagon centered at the $\Gamma$ point (Figs. 2 and  4(a)). The measured bands that are responsible for the linking features are marked by the white arrows in Fig. 4 (cuts 3 and 4), showing steep bands crossing the Fermi level.  Third, as shown in Fig. 4(b) (cuts 13 and 14), the two wings for the $\gamma$ (and $\gamma$$'$) bands close to the $K$ (and  $K'$) points form a hole-like band around the Brillouin zone boundary line in the second zone.   In addition to these three major features, there are other bands that also contribute to the Fermi surface in the second Brillouin zone. All these observations suggest the multiband nature of the PdTe$_2$ system at low energy which should be considered in understanding its superconductivity and related properties. Our present work provides for the first time detailed information on the complex electronic structure of the PdTe$_2$ system.

The $T_c$ enhancement in Cu$_{0.05}$PdTe$_2$ ($T_c=2.4$ K) compared to its parent compound PdTe$_2$ ($T_c=1.7$ K) prompted us to investigate the change in the electronic structures induced by the Cu intercalation. Figure 5 presents the direct electronic structure comparison between the intercalated Cu$_{0.05}$PdTe$_2$ (Figs. 5(d)--5(f), 5(i), 5(j)) and undoped PdTe$_2$ (Figs. 5(a)--5(c), 5(g), 5(h)) systems measured at $\sim20$ K using both the 21.218 eV and 6.994 eV photon energies.  Like the overall similarity of the Fermi surface shown in Fig. 2, the doped and undoped samples share similar band structures to a great detail for the measurements carried out with the same photon energy (Fig. 5). On the other hand, the measured band structure shows a big difference for the same momentum cut between the two different photon energies (compare Figs. 5(a), 5(c), 5(d), 5(f) with Figs. 5(g), 5(h), 5(i), 5(j)). While only a weak electron-like $\alpha$ band is observed near the $\Gamma$ point near the Fermi level in the 21.2 eV measurements (Figs. 3(b) and 3(c)),  both a hole-like narrow band, denoted as h-band that crosses the Fermi level, and an electron-like band denoted as e-band are observed in the 6.994 eV laser ARPES measurements (Figs. 5(g)--5(j)). We notice that, the electron-like e-band here has an energy range from $-0.46$ eV to 0 eV, which is consistent with the $\alpha$ band observed in the helium-lamp-based data. However they show a slight difference in the momentum range. A careful examination indicates that the e-band in the laser measurement has a momentum range from $-0.2$ {\AA}$^{-1}$ to $0.2$ {\AA}$^{-1}$ at the Fermi level in PdTe$_2$, while it is from $-0.15$ {\AA}$^{-1}$ to $0.15$ {\AA}$^{-1}$ for the $\alpha$ band at the Fermi level in the helium lamp measurement. Considering the different momentum resolutions of the two photon energies and the  photoemission matrix-element effect, we believe that the e-band observed by the laser ARPES is most likely the $\alpha$ band we identified in the helium lamp measurement.  But the hole-like h-band in the laser ARPES measurement is totally absent in the helium lamp measurements.  Besides the h-band, there are two other hole-like bands centered at the $\Gamma$ point in the laser measurement at higher binding energies (0.6--1.1 eV) which are not observed in the helium lamp experiment.  These differences reflect strong k$_z$ dependence of the three-dimensional electronic structure for PdTe$_2$, as we will see in the following band structure calculations.

The band structure change induced by Cu-intercalation can be revealed by comparing the core level energy shifts. Figure 5(l) shows the energy distribution curves (EDCs) at the $\Gamma$ point taken from both the $\Gamma$--M and $\Gamma$--K momentum cuts for PdTe$_2$ and Cu$_{0.05}$PdTe$_2$ samples measured by using 21.2 eV photon energy. Although there is a slight relative intensity variation of the observed peaks between these two samples, overall in the large energy range, there is no obvious peak position shift within our measurement precision.  On the other hand, using the high resolution vacuum ultra-violet laser ARPES, we can see obvious relative energy shifts in the EDCs at the $\Gamma$ point between the doped and undoped PdTe$_2$ systems (Fig. 5(m)). The features in the covered energy range for Cu$_{0.05}$PdTe$_2$ exhibit a shift of $\sim$16 meV towards the high binding energy. This shift can also be corroborated  by the change of the distance between the two Fermi momenta k$_F$ of the hole-like h-bands at the Fermi level (Fig. 5(n));  it shows a clear shrink ($\sim$0.03 {\AA}$^{-1}$) for the Cu$_{0.05}$PdTe$_2$  sample, consistent with the chemical potential shift towards higher binding energy. These observations indicate that Cu-intercalation in PdTe$_2$ induces electrons into the system, which may be related to the enhancement of $T_c$ in Cu$_{0.05}$PdTe$_2$.  We note that the energy shifts of the bands are observed only near the $\Gamma$ point due to the limited momentum space covered by the laser ARPES, more work needs to be done to examine the change  on the $\beta$ ($\beta$$'$) bands since they are the dominant features in the measured Fermi surface.

In order to understand the electronic structure of PdTe$_2$ and have a direct comparison with the experiment, band structure calculations were carried out. Figure 6 shows the calculated  band structure (Figs. 6(a) and 6(b)), the density of state (DOS) (Figs. 6(c) and 6(d)), and the three-dimensional Fermi surface (Figs. 6(e) and 6(f)), without and with considering the spin--orbit coupling.  Our calculations overall agree with the previous report.\cite{GYGuo}  First, the electronic structure of PdTe$_2$ shows strong three-dimensionality, as manifested from the different band structures on the MGK  and LAH planes (Figs. 6(a) and 6(b)), and the three-dimensional Fermi surface topology seen in Figs. 6(e) and 6(f). The strong k$_z$ dependence of the constant energy contours and the band structure in Fig. 7 also support this picture.   Second, the low-energy electronic states are mainly contributed by the Pd 4d states and Te 5p states, as seen from the calculated DOS in Figs. 6(c) and 6(d).  At high binding energy above 2 eV, the Pd 4d state contributes most of the spectral weight, while around the Fermi level,  the Te 5p state and Pd 4d state contribute nearly equally (Figs. 6(c) and 6(d)). It is interesting to note that  the Fermi level $E_F$ is located near the valley of the overall density of states, indicating the existence of a low carrier density and that it is relatively easy to tune $T_c$.\cite{Myron} The detailed spectral weight contributions of the Pd 4d and Te 5p orbitals are shown in Figs. 6(a2), 6(a3) and Figs. 6(b2), 6(b3) for each of the hybridized bands.  Third, the spin--orbit coupling plays an important role in the electronic structure of PdTe$_2$. While there are three bands that cross the Fermi level in the calculated band structure without considering the spin--orbit coupling (Fig. 6(a)), four band crossings are present in the calculated results with the spin--orbit coupling (Fig. 6(b)).  The spin--orbit coupling also induces band splitting of several bands, as seen from Figs. 6(a) and 6(b). These results suggest that the spin--orbit coupling in PdTe$_2$ has an obvious effect on the electronic structure. Our simulation on the spin--orbit coupling strength gives a value of $\sim$0.5 eV for PdTe$_2$.

Figure 7 shows a comparison between our experimental results and the band structure calculations. Since the ARPES band structure measurement at a given photon energy provides only one line momentum cut in the three-dimensional
 momentum space, and the measured Fermi surface only represents a (curved) surface cut through the three-dimensional Fermi surface, it is important to locate k$_z$ of the ARPES measurements for the direct comparison.  In the
 band structure calculations,  we have calculated the Fermi surface and band structures along high symmetry momentum cuts at various k$_z$.  Figure 7b shows the calculated Fermi surface at four different k$_z$ (k$_z$=0,1/3,
  2/3, and 1) from $\Gamma$ to A point with  equal k$_z$ increment, the value of k$_z$ is in units of $\pi/c$.  For $k_z=0$,  there is an enclosed large polyhedron hole-like Fermi surface sheet surrounding the $\Gamma$ point
   (Fig. 7(b1)), and two pairs of irregular pockets appear at the $K$ points.  The Fermi surface topology varies strongly with k$_z$ due to three-dimensionality of the electronic structure. For k$_z$=1, one sees only six small
   electron pockets in the first Brillouin zone. The corresponding band structure also shows a strong k$_z$ dependence as seen in Fig. 7(d).

Direct comparison between the measured and the calculated Fermi surfaces and band structures indicates that, our measured results fit closest to the k$_z$=1 case. The dominant feature, the observation of electron-like
bright spots $\beta(\beta')$ in the measured Fermi surface (Fig. 7(a)), shows the best fit to the calculated Fermi surface topology near k$_z$=1 (Fig. 7(b4)). This provides a way to estimate and check on the location of
k$_z$ in our ARPES measurements using the helium lamp with a photon energy of 21.2 eV and the VUV laser with a photon energy of 6.994 eV.  The momentum perpendicular to the sample surface during the photoemission process
is given by k$_z$=0.512$\times$$\sqrt{(h\nu-W)\cos^{2}\theta+V_{0}}$ in units of \AA$^{-1}$.\cite{Ootsuki2014PRB} Considering k$_z$ to be 1 for 21.2 eV photon energy,  the inner potential V$_0$ can be roughly estimated to
be 18.8 eV. This value is similar to that measured in other related transition metal dichalcogenides such as WTe$_2$ with  $V_0=14.5$ eV\cite{YZhang} and IrTe$_2$ with $V_0=14$ eV.\cite{Ootsuki2014PRB}  This value can be
further checked by our laser ARPES measurements, where k$_z$=3.88$\pi$/c is obtained with photon energy $h\nu=6.994$ eV. This is close to the k$_z$=0 $\Gamma$MK plane, and indeed the observation of hole-like bands in the
laser-ARPES measurements (Figs. 5(g)--5(j)) is consistent with the calculated band structure at k$_z$=0 (Fig. 7(d1)). These results indicate that our selection of V$_0$ and k$_z$ can capture the major observations of our
measurements.

However, we note that, although there is a good correspondence on the major features, the deviations of our measurements from the band structure calculations are also obvious.  The most notable one is the simultaneous observation of multiple Fermi surface sheets, like the $\alpha$ band near $\Gamma$, the two flaps ($\gamma$ bands) around the K points, the $\delta$ bands extending out from the $\beta$ bright spots, and other features in the second Brillouin zone. Some of them are difficult to find correspondence in the calculated Fermi surface at different k$_z$ planes. The simultaneous observation of multiple features may originate from poor k$_z$ momentum resolution which is not known at the point. Further detailed photon energy-dependent ARPES measurements may help clarify this issue.

\section{Conclusion}
We report detailed electronic structures of PdTe$_2$ and Cu$_{0.05}$PdTe$_2$ samples studied by high resolution angle-resolved photoemission measurements as well as the first-principle band structure calculations. Multiple bands and complex Fermi surface topologies are revealed for these systems, with three singular bright spots identified in the measured Fermi surface at 21.2 eV photon energy. Both our experiment and band structure calculations indicate strong three-dimensionality of the electron structure and the importance of the spin--orbital coupling in PdTe$_2$. Our results indicate that Cu-intercalation in PdTe$_2$ introduces electrons into the system, which may be related to the T$_c$ enhancement in Cu$_{0.05}$PdTe$_2$. While the band structure calculations can capture some major features observed in our experiment, obvious discrepancy exists.  We believe our present comprehensive and systematic work on PdTe$_2$  and Cu$_{0.05}$PdTe$_2$ systems will lay a good foundation for further research on transition metal dichalcogenides and the related superconductors. Further photon energy dependent ARPES measurements are needed to reveal the three-dimensional electronic structure of PdTe$_2$ and resolve the deviation between the band structure calculations and the experiment. Further theoretical study is needed to consider the effect of the surface state and investigate the related topological property in the PdTe$_2$ system. Extremely-low temperature and high resolution ARPES experiments are required to study the electronic behaviors and superconductivity mechanism in the superconducting state of PdTe$_2$ related systems.

\vspace{3mm}

\noindent {\bf Acknowledgement} XJZ thanks financial support from the NSFC (11190022), the MOST of China (973 program No: 2011CB921703 and 2011CBA00110), and  the Strategic Priority Research Program (B) of the Chinese Academy of Sciences (Grant No. XDB07020300).

\vspace{3mm}

\vspace{3mm}

\noindent {\bf Author Contributions}\\
X.J.Z. and Y.L. proposed and designed the research. C.T.L. contributed in sample preparation. Y.L., L.Y.,C.H., D.F.L.,Y.Y.P., Z.J.X., J.F.H., C.Y.C., Y.F., H.M.Y., X.L., L.Z., S.L.H.,G.D.L., J.Z., C.T.C., Z.Y.X. and X.J.Z. contributed to the development and maintenance of Laser-ARPES system. Y.L. carried out the ARPES experiment.  Y.L., L.Y. and X.J.Z. analyzed the data. Y.L., J.Z.Z., H.M.W., X.D. and Z.F. performed band structure calculations.  Y.L., L.Y.  and X.J.Z. wrote the paper and all authors participated in discussion and comment on the paper.

\vspace{3mm}

\noindent {\bf\large Additional information}\\

\noindent{\bf Competing financial interests:} The authors declare no competing financial interests.

\newpage

\begin{figure*}[tbp]
\begin{center}
\includegraphics[width=1.0\columnwidth,angle=0]{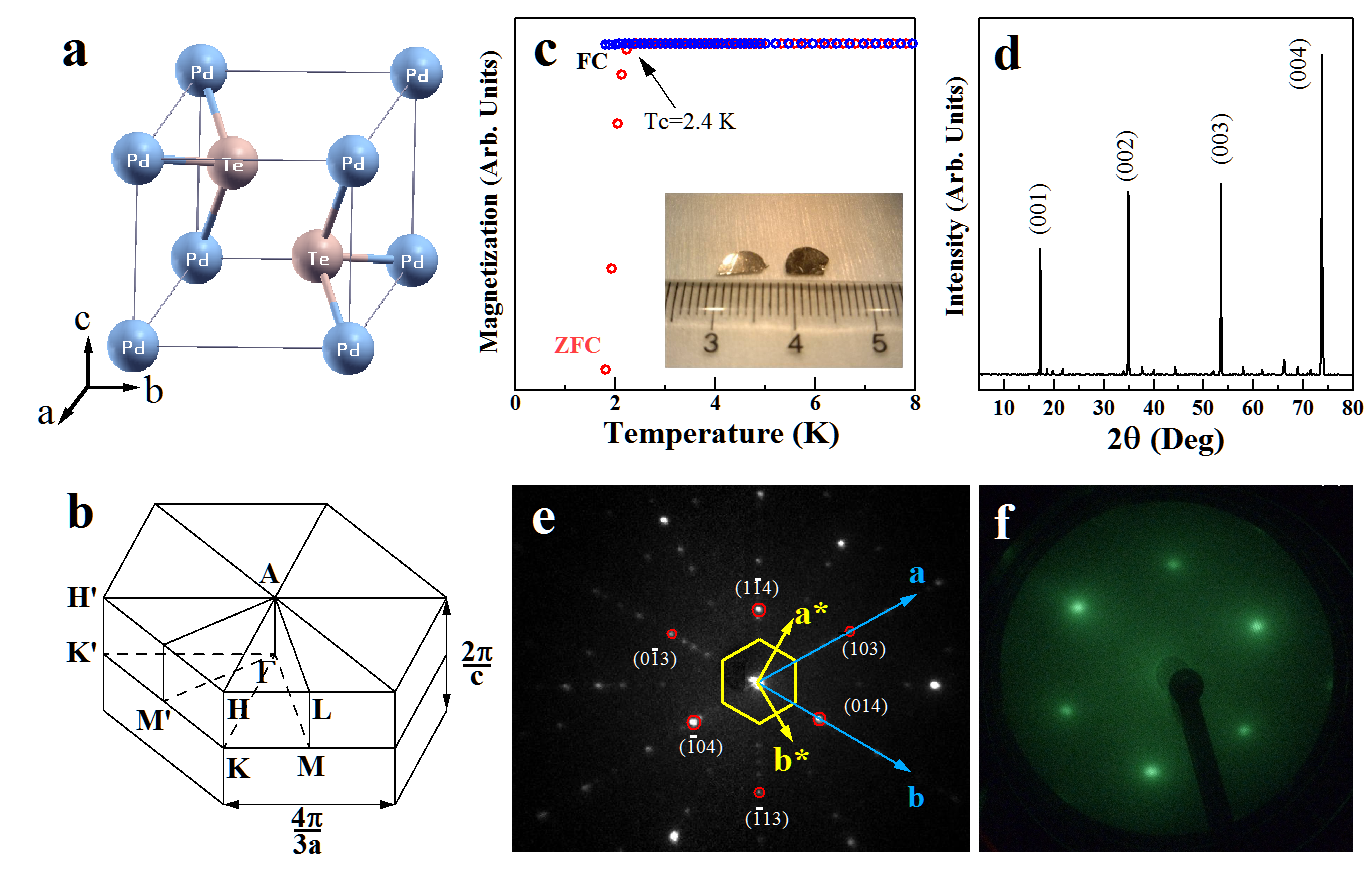}
\end{center}
\caption{(color online) Crystal structure and Brillouin zone of PdTe$_2$ and sample characterization of Cu$_{0.05}$PdTe$_2$ single crystal. (a) Crystal structure of PdTe$_2$ with a polymeric CdI$_2$-type structure (Space group: P-$\bar{3}$m1). (b) The corresponding hexagonal prism Brillouin zone of PdTe$_2$. (c) Magnetization measurement of Cu$_{0.05}$PdTe$_2$ single crystal with a magnetic field of 1 Oe. FC (blue circles) refers to the field-cooled measurement, while ZFC (red circles) refers to the zero-field-cooled measurement. The bottom-right inset shows a photograph of the plate-like Cu$_{0.05}$PdTe$_2$ single crystal with a shining surface and  a dimension of 5 mm$\times$3 mm$\times$0.2 mm. (d) The x-ray diffraction pattern of a naturally cleaved surface of Cu$_{0.05}$PdTe$_2$ single crystal, showing that it is the basal plane. (e) Laue diffraction pattern of the Cu$_{0.05}$PdTe$_2$ single crystal at room temperature. The six red circles mark the six spots we used for indexing search by Laue simulation.  The blue axes are the determined lattice vectors $\bm{ a}$ and $\bm{ b}$ in the real space from the Laue indexing. The yellow axes are the determined reciprocal lattice vectors $\bm{ a}^{\ast}$ and $\bm{ b}^{\ast}$. The yellow hexagon is the first Brillouin zone determined by the reciprocal lattice vectors. (f) The low-energy electron diffraction (LEED) pattern of a freshly cleaved Cu$_{0.05}$PdTe$_2$ surface taken with an electron energy of 70 eV at 25 K.}
\end{figure*}



\begin{figure*}[tbp]
\begin{center}
\includegraphics[width=1.0\columnwidth,angle=0]{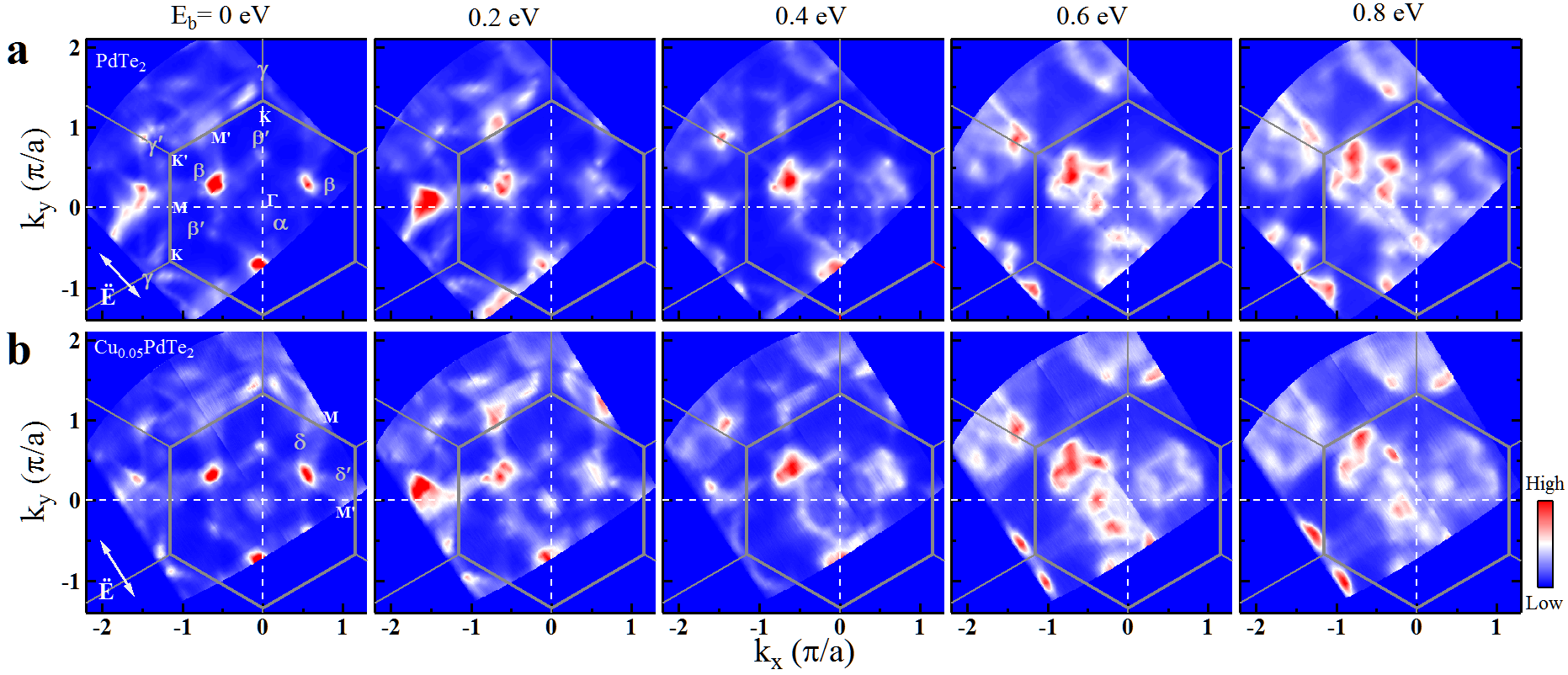}
\end{center}
\caption{(color online) Constant energy contours of PdTe$_2$ and Cu$_{0.05}$PdTe$_2$ measured at 20 K at different binding energies. (a) Constant energy contours of PdTe$_2$ at different binding energies of E$_b$ $\sim$ 0 eV, 0.2 eV, 0.4 eV, 0.6 eV, 0.8 eV. (b) Constant energy contours of Cu$_{0.05}$PdTe$_2$ at different binding energies of E$_b$ $\sim$ 0 eV, 0.2 eV, 0.4 eV, 0.6 eV, 0.8 eV.}
\end{figure*}

\begin{figure*}[tbp]
\includegraphics[width=1.0\columnwidth,angle=0]{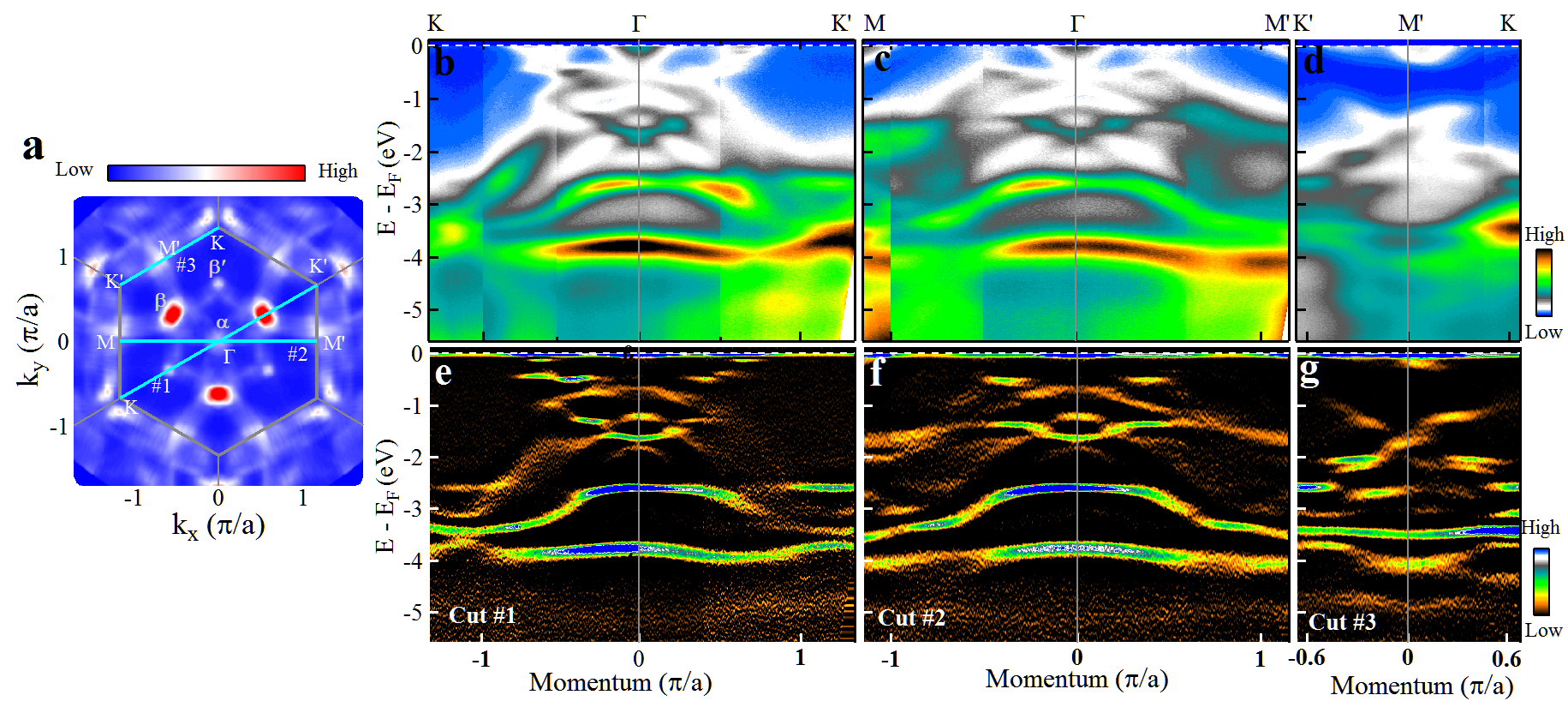}
\begin{center}
\caption{(color online) Band structures of Cu$_{0.05}$PdTe$_2$ measured along three high-symmetry directions using the helium lamp ARPES. (a) Symmetrized Fermi surface of Cu$_{0.05}$PdTe$_2$. The locations of the three momentum cuts are marked as blue lines \#1, \#2, and \#3.  (b)--(d) Band structures measured along three high symmetry directions: (b) K-$\Gamma$-K$'$ (cut 1), (c) M-$\Gamma$-M$'$ (cut 2), and (d) K$'$-M$'$-K (cut 3). Their corresponding energy-second-derivative images are shown in panels (e)--(g), respectively.
}
\end{center}
\end{figure*}

\begin{figure*}[tbp]
\begin{center}
\includegraphics[width=1.0\columnwidth,angle=0]{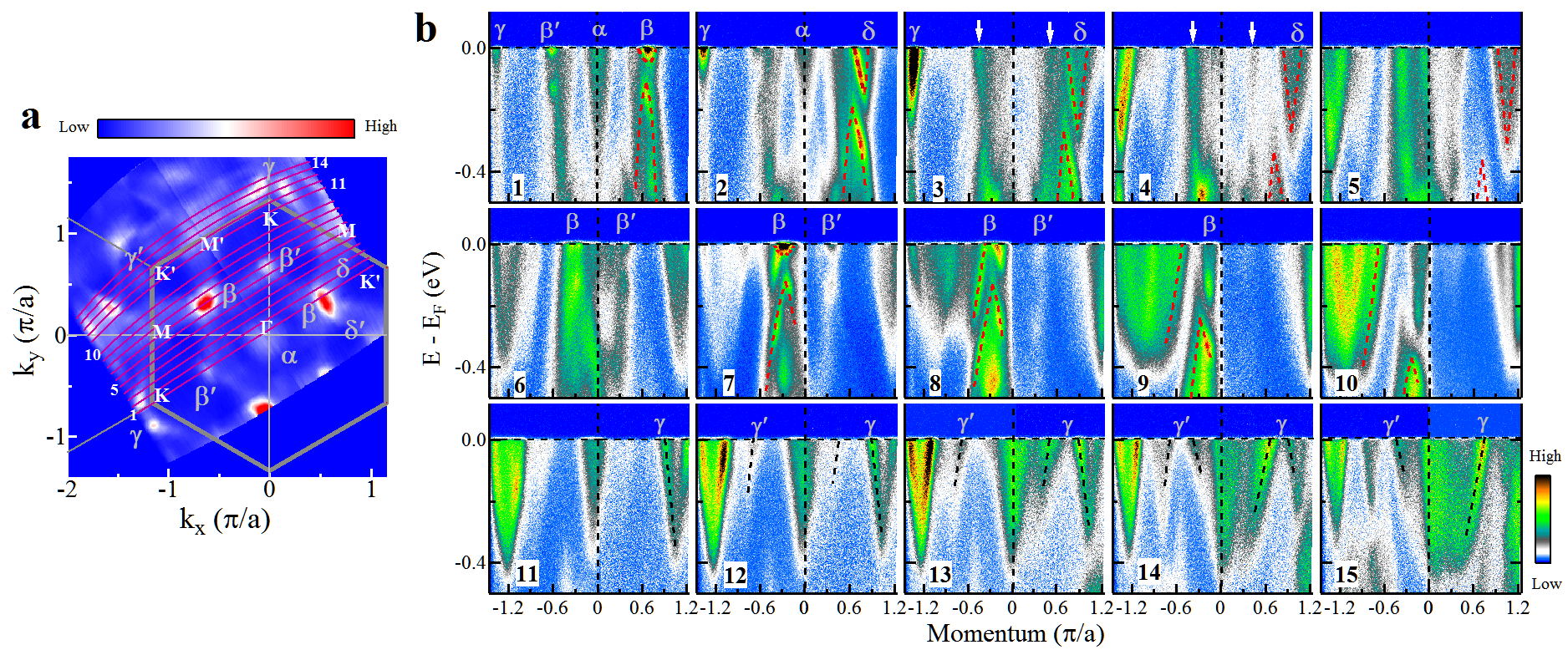}
\end{center}
\caption{(color online) Momentum dependence of the band structures of Cu$_{0.05}$PdTe$_2$ measured by using the helium lamp ARPES. (a) Fermi surface of Cu$_{0.05}$PdTe$_2$ with the major Fermi surface features $\alpha$, $\beta$ ($\beta$$'$),  $\gamma$ ($\gamma$$'$), and $\delta$ ($\delta$$'$) marked. The locations of the momentum cuts are marked by red lines and represented by numbers 1--15.  (b) Band structures along the momentum cuts 1--15. The corresponding bands which contribute to $\alpha$, $\beta$ ($\beta$$'$),  $\gamma$ ($\gamma$$'$), and  $\delta$ ($\delta$$'$) Fermi surface sheets are noted. The red dashed lines indicating the $\beta$ bands and the black dashed lines indicating the $\gamma$ bands are guide to eyes. }
\end{figure*}

\begin{figure*}[tbp]
\begin{center}
\includegraphics[width=1.0\columnwidth,angle=0]{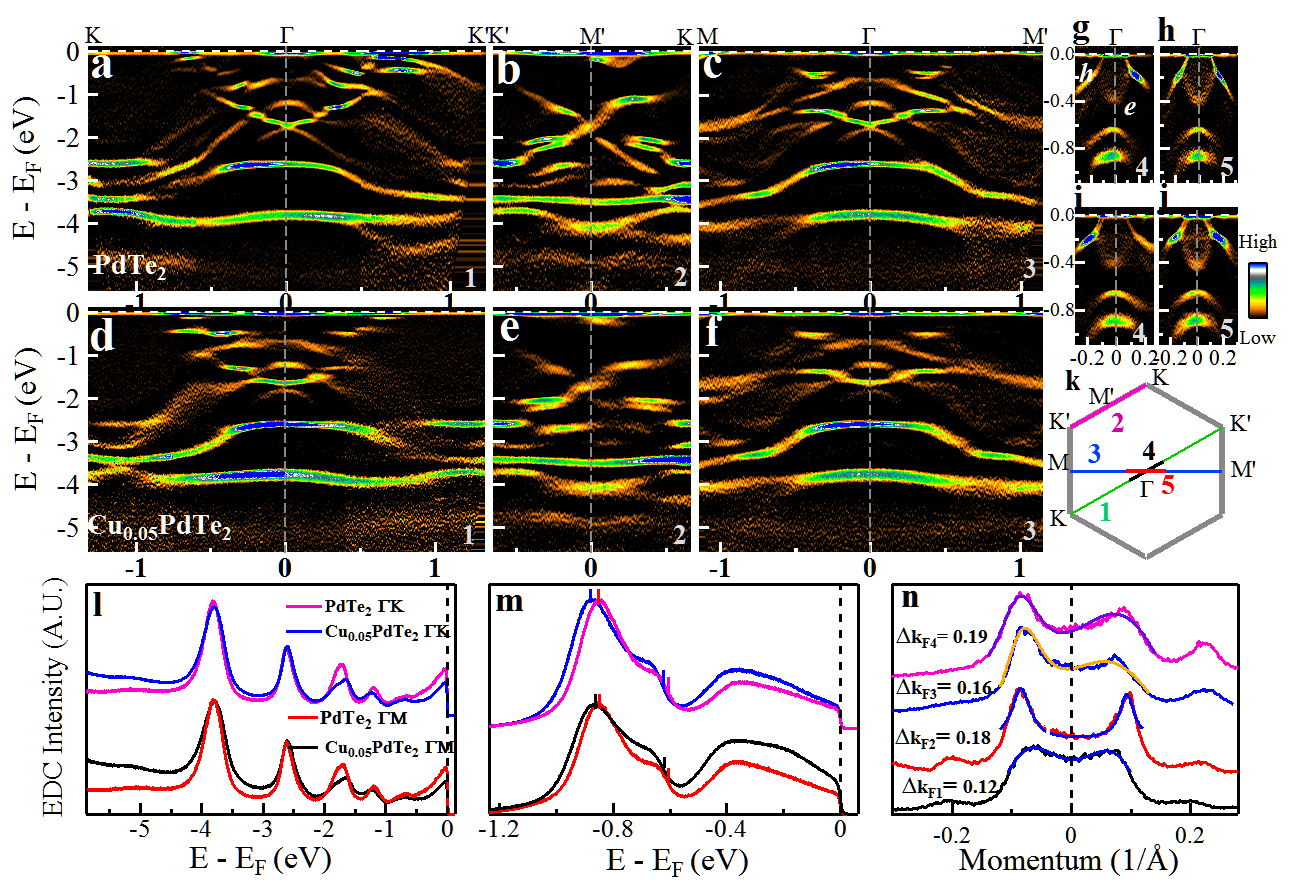}
\end{center}
\caption{(color online) Band structure comparison between PdTe$_2$ and Cu$_{0.05}$PdTe$_2$ by using hellium I (21.218 eV) and laser (6.994 eV) as the photon source.  (a)--(c) Band structures of PdTe$_2$ measured by He I along three high symmetry cuts 1--3.  (d)--(f) Band structures of Cu$_{0.05}$PdTe$_2$ along three high symmetry cuts 1--3.  (g) and (h) Band structure of PdTe$_2$ measured by laser along two high symmetry cuts 4 and 5. (i) and (j) Band structure of Cu$_{0.05}$PdTe$_2$ measured along two high symmetry cuts 4 and 5.  The locations of the five momentum cuts 1--5 are shown in panel (k).  To highlight the measured bands, all the images shown here are the second-derivative images of the original data with respect to the energy. Panel (l) shows the energy distribution curves (EDCs) of Cu$_{0.05}$PdTe$_2$ and PdTe$_2$ at the $\Gamma$ point in panels (a)--(c) and (f) from the helium I measurements. Panel (m) shows the extracted EDCs of Cu$_{0.05}$PdTe$_2$ and PdTe$_2$ at the $\Gamma$ point in panels (g)--(j) from the laser-ARPES measurement. The vertical bars indicate the peak position in EDCs. Panel (n) shows the momentum distribution curves (MDCs) at the Fermi level of panels (g)--(j). The fitted MDC lines are put on top of the measured data and the obtained Fermi momentum difference is marked.}
\end{figure*}

\begin{figure*}[tbp]
\begin{center}
\includegraphics[width=1.0\columnwidth,angle=0]{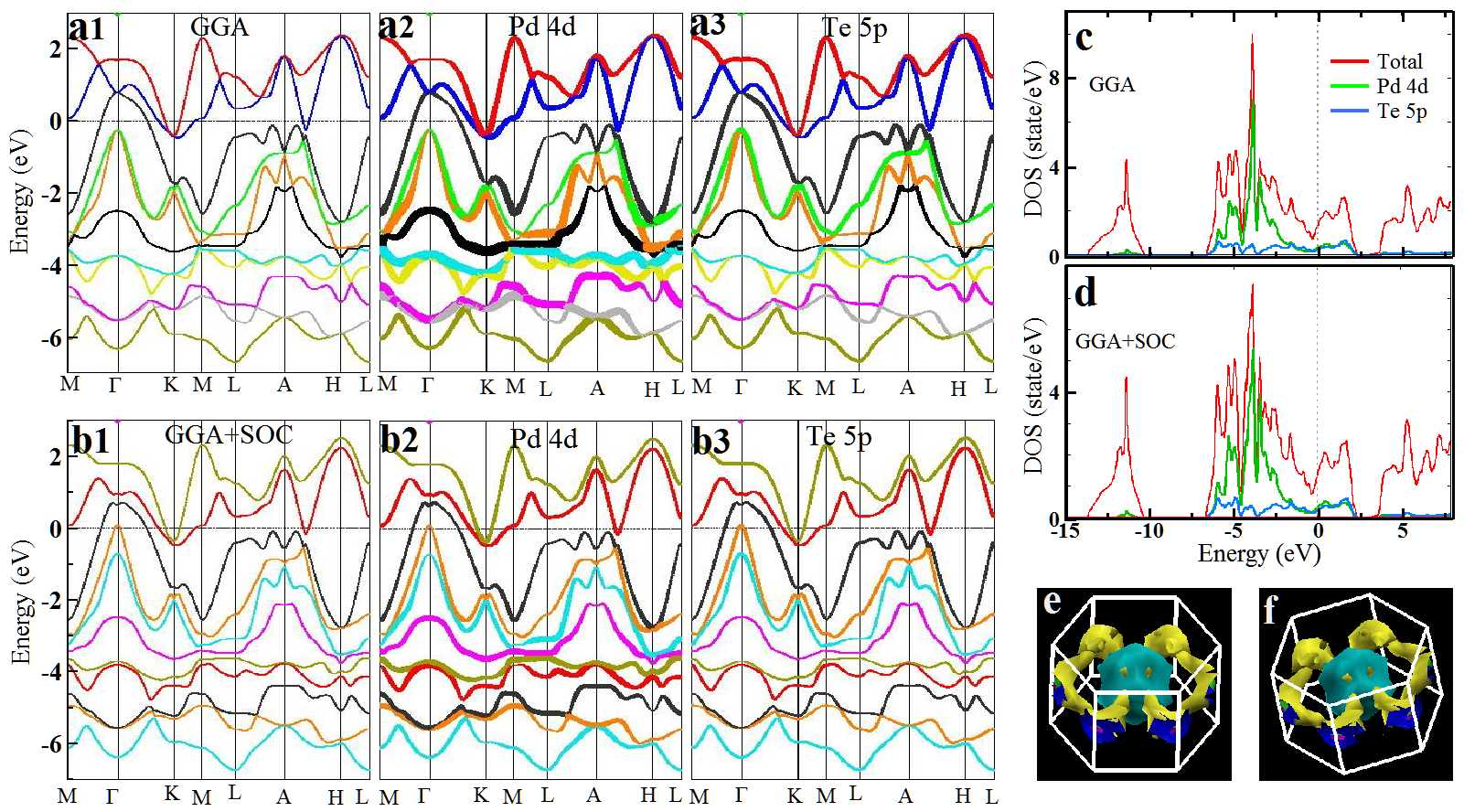}
\end{center}
\caption{(color online) Calculated band structure, density of states, and Fermi surface of PdTe$_2$ with and without considering the spin--orbit coupling. (a1)--(a3) Calculated band structure of PdTe$_2$ without the spin--orbit coupling. (b1)-(b3) Calculated band structure of PdTe$_2$ with the spin--orbit coupling.  The band character calculations with Pd 4d states ((a2), (b2)) and Te 5p states ((a3), (b3)) highlighted are shown, respectively. Each state is drawn by a dot whose radius is proportional to the specified character of the state. (c) and (d) Calculated density of state without and with the spin--orbit coupling.   (e) and (f) Calculated three-dimensional Fermi surface of PdTe$_2$ without and with the spin--orbit coupling.
}
\end{figure*}

\begin{figure*}[tbp]
\begin{center}
\includegraphics[width=1.0\columnwidth,angle=0]{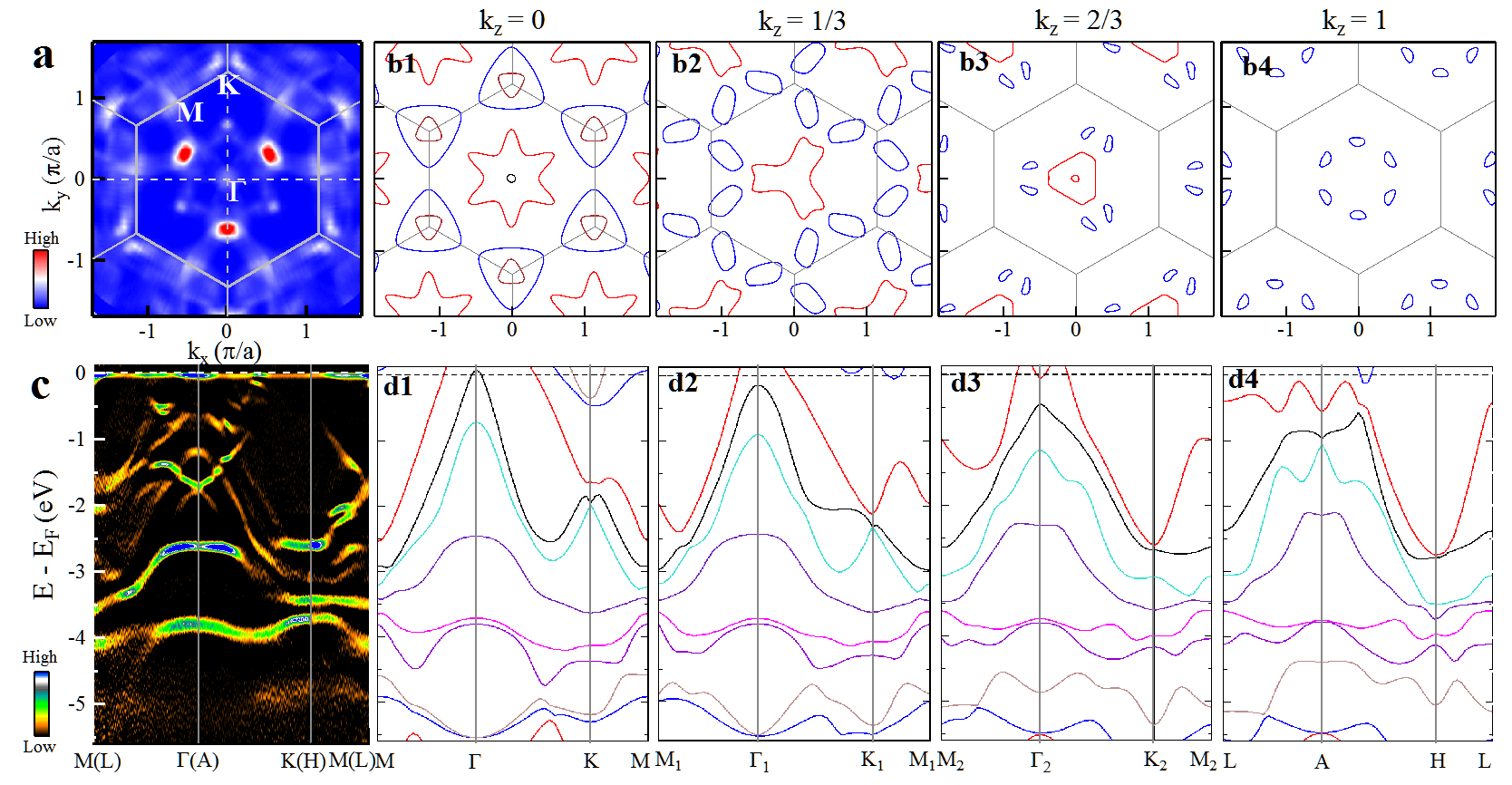}
\end{center}
\caption{(color online) Calculated k$_z$ dependence of the Fermi surface and band structures of PdTe$_2$ with the spin--orbit coupling and the comparison with the measurements. (a) Fermi surface of PdTe$_2$ measured by using the helium I ARPES. (b1)--(b4) Calculated Fermi surface at  k$_z=0$, 1/3, 2/3, and 1, respectively. The value of k$_z$ is in units of $\pi$/c. (c) Band structure measured along high symmetry directions  in the helium I ARPES experiment. To highlight the measured bands, we show the energy-second-derivative images of the original data. (d1)--(d4) Calculated band structures at  $k_z=0$, 1/3, 2/3, and 1, respectively.}
\end{figure*}

\end{document}